\documentclass[aps,twocolumn,floatfix]{revtex4-1}

\usepackage[dvips]{graphicx}

\begin{document}

\title{Fate of topological states in incommensurate generalized
  Aubry-Andr\'e models}

\author{J. C. C. Cestari} 

\affiliation{Universidade Federal do Rio Grande do Sul, C.P. 15051,
  91501-970 Porto Alegre, Brazil}

\author{A. Foerster}

\affiliation{Universidade Federal do Rio Grande do Sul, C.P. 15051,
  91501-970 Porto Alegre, Brazil}

\author{M. A. Gusm\~ao} 

\affiliation{Universidade Federal do Rio Grande do Sul, C.P. 15051,
  91501-970 Porto Alegre, Brazil}

\begin{abstract}
  We study one-dimensional optical lattices described by generalized
  Aubry-Andr\'e models that include both commensurate and
  incommensurate modulations of the hopping amplitude. This brings
  together two interesting features of this class of systems: Anderson
  localization and the existence of topological edge states. We follow
  changes of the single-particle energy spectrum induced by variations
  of the system parameters, with focus on the survival of topological
  states in the localized regime.
\end{abstract}


\maketitle

\section{Introduction}
In recent years, rapid progress in techniques for creating
ultracold-atom systems in laboratory allowed the experimental
realization of many interesting models originally proposed to study
specific properties of real solids. For instance, the construction of
bichromatic lattices with incommensurate potentials led to observation
\cite{Billy08,Roati08,pugatch2} of the Anderson-localization
transition \cite{Anderson} in one dimension, which cannot happen for
true disorder. Such a transition has received the attention of
theoreticians for many years
\cite{AndLocSarma01,AndLocSarma02,AndLocSarma03,AndLocSarma04,
  cestari10,cestari12}. In this context, the standard theoretical
approach utilizes the Aubry-Andr\'e (AA) model \cite{AA}, intimately
related to the Harper-Hofstadter (HH) model \cite{harper,Hofstadter}
for electrons in a two-dimensional (2D) lattice in the presence of a
perpendicular magnetic field. The latter is mapped onto a
one-dimensional (1D) system with a modulating potential superimposed
to the lattice, its period ($1/\beta$) being determined by the
magnetic-field intensity. Thus, the relative periodicities between
modulating potential and lattice can be tuned in principle to any
ratio. The energy spectrum for varying $\beta$ appears as the famous
Hofstadter butterfly \cite{Hofstadter}. This link between the 2D HH
and 1D AA models has also been explored from the point of view of
topological properties \cite{topo1, KrausZilb01,KrausZilb02}. This
revealed connections with seemingly unrelated systems, such as
topological insulators \cite{topo_insul} and superconductors
\cite{majorana}, as well as the Quantum Hall Effect (QHE)
\cite{QHE01,QHE02,QHE03}.

Lately, extensions of the AA model have been proposed
\cite{Liuetal2015,topolSarma} including periodic modulations of the
nearest-neighbor hopping amplitude.  An incommensurate hopping
modulation leads to Anderson-like localization \cite{Liuetal2015},
thus mimicking disorder, as the diagonal AA potential, while a
commensurate modulation brings up new features, like the appearance of
zero-energy topological edge states \cite{topolSarma}.  Here we
combine commensurate and incommensurate off-diagonal modulations,
which turns out to be nontrivial from the point of view of
topological properties. Indeed, we find that topological edge states
are robust against an incommensurate perturbation, surviving the
localization transition in a certain range of parameters. This result
opens new perspectives for the investigation of the interplay between
topology and disorder.

A generalized Aubry-Andr\'e model, including commensurate and
incommensurate hopping modulations as well as a diagonal
incommensurate potential, may be described by the Hamiltonian
\begin{equation} 
\label{eq:HAA} 
H = -t \sum_{i} \left[ (1 + \lambda_i + \delta_i)  (a^{\dag}_{i+1}
a_i^{} + \mathrm{H.c.})  + \varepsilon_i a^{\dag}_i a_i \right]\!\!,
\end{equation}
where 
\begin{eqnarray} 
\label{eq:params}
\lambda_i &=& \lambda \cos(2\pi b i + \varphi)\,, \quad
\delta_i = \delta \cos(2\pi \beta i + \varphi_\delta)\,, \nonumber \\
\varepsilon_i &=& \Delta \cos(2\pi\beta i + \varphi_\Delta) 
\end{eqnarray}
are commensurate and incommensurate hopping modulations, and the
diagonal AA potential, respectively; $i$ assumes integer values
labeling lattice sites; $t$ represents the hopping (or tunneling)
amplitude; the creation and annihilation operators $a^{\dag}_i$ and
$a^{}_i$ can be bosonic or fermionic (differences being in the nature
of many-body states).  The phases in the three periodic terms are
possibly all different. The inverse wavelengths of commensurate and
incommensurate modulations are respectively denoted as $b$ and
$\beta$. We will mostly focus on the case $b=1/2$, and we fix $\beta=
(1 + \sqrt{5}\,)/2$, the golden ratio.  Without off-diagonal
modulation ($\lambda=\delta=0$), one recovers the usual AA model. For
simplicity, we will refer to $\lambda$ as \emph{modulation amplitude},
and to $\Delta$ and $\delta$ as (respectively, diagonal and
off-diagonal) \emph{disorder strengths}, since the incommensurate
terms can be viewed as a kind of (non-random) disorder.

Our aim is to investigate how the disorder perturbations affect the
spectrum obtained in the commensurate case ($\Delta=\delta=0$), with
special attention to what happens to the topological states. We will
do this by exact diagonalization on finite lattices. Complementing a
direct visualization of the energy spectrum as it evolves under the
perturbation, a more detailed analysis of its changes will be done by
calculating the superfluid fraction \cite{roth}, and the ground-sate
\emph{fidelity} \cite{nielsen}. This latter quantity is known to
be a powerful tool to detect precursors of quantum phase transitions
(QPT's). Particularly for the kind of lattice models addressed here,
we have previously shown \cite{cestari10} that it is sensitive to
ground-state changes at the QPT critical parameters even for fairly
small systems.

\section{Localization}

As mentioned above, an Anderson-like localization transition occurs in
the usual AA model ($\lambda= \delta = 0$) for a critical
$\Delta_c=2\,t$ when $\beta$ is the golden ratio. In a Bose-Einstein
condensate, the localized phase is characterized by a null value of
the superfluid fraction, which is calculated by imposing twisted
periodic boundary conditions with a small twist angle $\theta$. The
superfluid fraction $f_s$ is then proportional to the energy
difference between twisted and non-twisted ground states divided by
$\theta^2$. For finite lattices, it is necessary to utilize a
golden-ratio approximation as the quotient between two consecutive
Fibonacci numbers, one of which is the number of lattice sites
\cite{kohmoto1}. On the other hand, the ground-state fidelity, in
this case defined as the scalar product between two ground-state
vectors corresponding to slightly different values of $\Delta$, is
able to detect the transition as a sharp minimum at $\Delta_c$, both
with periodic or open boundary conditions. 

\begin{figure}
   \includegraphics[width=7.5cm]{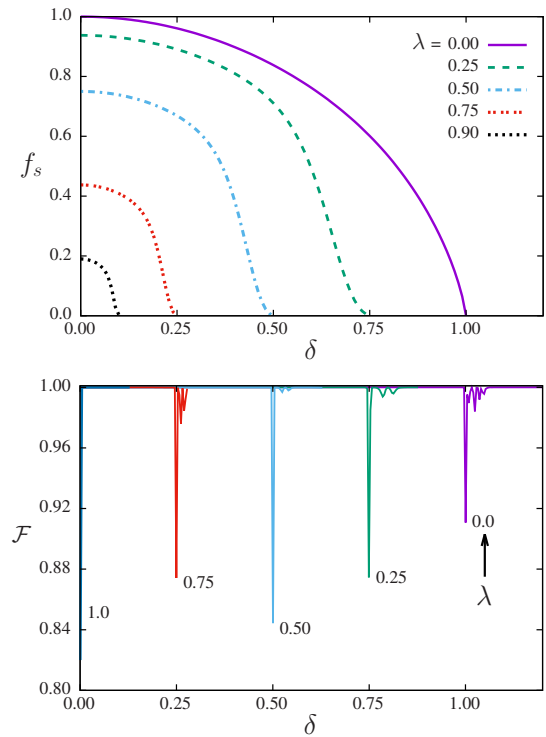} 
   \caption{Superfluid fraction (top) and ground-state fidelity
     (bottom) as functions of disorder strength $\delta$ for the
     indicated values of modulation amplitude $\lambda$. Notice the
     coincidence of critical points.}
  \label{fig:sf_fid}
\end{figure}

Anderson localization also occurs in generalized AA models with
combined diagonal and off-diagonal disorder \cite{Liuetal2015}. Here,
we focus on purely off-diagonal disorder [$\Delta=0, \delta\ne 0$ in
Eqs.~(\ref{eq:HAA})-(\ref{eq:params})] but in the presence of
commensurate hopping modulation ($\lambda\ne 0$), in order to have
zero-energy topological states. It turns out that the critical
disorder strength $\delta_c$ depends on the modulation amplitude
$\lambda$. The top panel of Fig.~\ref{fig:sf_fid} shows the superfluid
fraction $f_s$ as a function of $\delta$ for $b = 1/2$ and different
values of $\lambda$.  One can clearly see critical values of $\delta$
at which $f_s$ drops to zero, indicating localization. In the bottom
panel of Fig.~\ref{fig:sf_fid} we plot the fidelity between two
ground-states differing by a small variation in $\delta$. This
fidelity has pronounced minima exactly at the values ​​of $\delta$ for
which the superfluid fraction vanishes, consistent with their
identification as critical values ​​for a localization transition.

The superfluid-fraction curves in Fig.~\ref{fig:sf_fid} were obtained
with periodic boundary conditions on lattices of 144 sites, with
$\beta = 233/144$, a rational approximant of the golden ratio. On the
other hand, fidelity values shown in the same figure were calculated
with open boundary conditions for chains of 200 sites. 
The coincidence of $\delta_c$ values ​​is remarkable. Other lattice
sizes were checked with essentially coincident results.

A superfluid fraction in principle implies a bosonic system, while
related problems, like topological insulators and superconductors
involve fermions. However, the single-particle energy spectrum is the
same, and our focus is on the non-interacting limit. A fundamental
difference would be the relevance of the Fermi level rather than the
lowest-energy state, but localization occurs for all states in one
dimension. In practice, the superfluid fraction is used here only to
indicate the presence of extended or localized states. It should also
be noticed that $f_s$ is proportional to the \emph{helicity modulus}
\cite{fisherpai}, which is more general, an can be viewed as a measure
of wavefunction coherence across the system.

The values of $\delta_c$ that we obtained obey a simple linear
relation, $\delta_c(\lambda) = 1 - \lambda$. The maximum value of
$f_s$ is also strongly dependent on $\lambda$, as seen in
Fig.~\ref{fig:sf_fid}. In particular, the curves $f_s (\delta)$ tend
to a single point ($f_s = 0$, $\delta = 0$) for $\lambda = 1$. The
kind of localization that occurs for $\lambda = 1$ when $\delta =0$
can be understood as a cancellation of the uniform hopping term with
the modulated one. Since $\cos(\pi i) = \pm 1$ for odd/even $i$, the
net hopping amplitude $t(1 + \lambda_i)$ alternates between $2t$ and
$0$, so that the 1D lattice breaks down into isolated dimers, and the
localization becomes \emph{trivial}. If we then turn on the
incommensurate hopping term, we find that the superfluid fraction
remains zero, since this essentially random connection is not capable
of building up extended states. All these results concerning
localization were obtained for zero phases ($\varphi_\delta = \varphi
= 0$). The effect of non-zero phases will be discussed in the
following.

\section{Topological states}

\begin{figure*}
  \includegraphics[width=5.8cm]{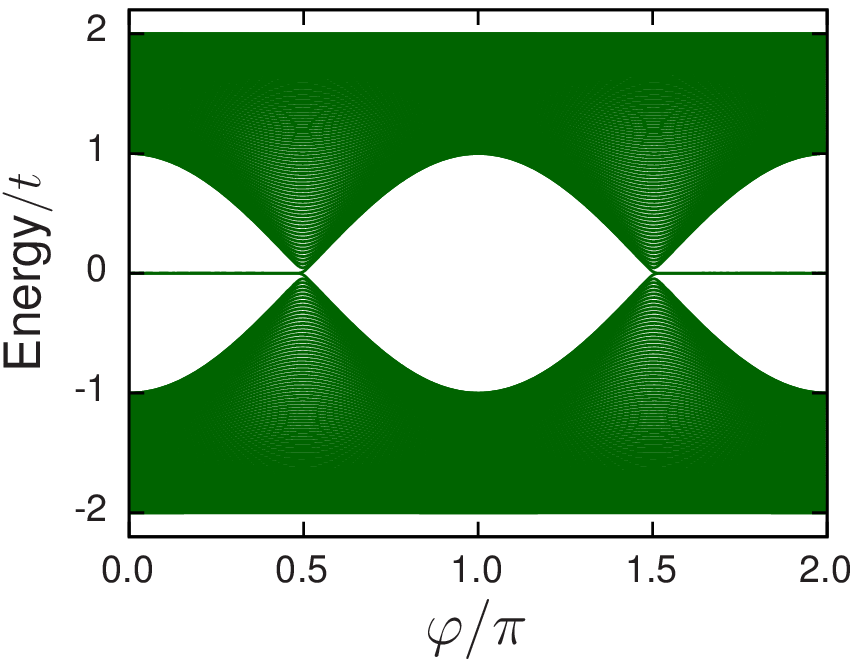} 
  \includegraphics[width=5.8cm]{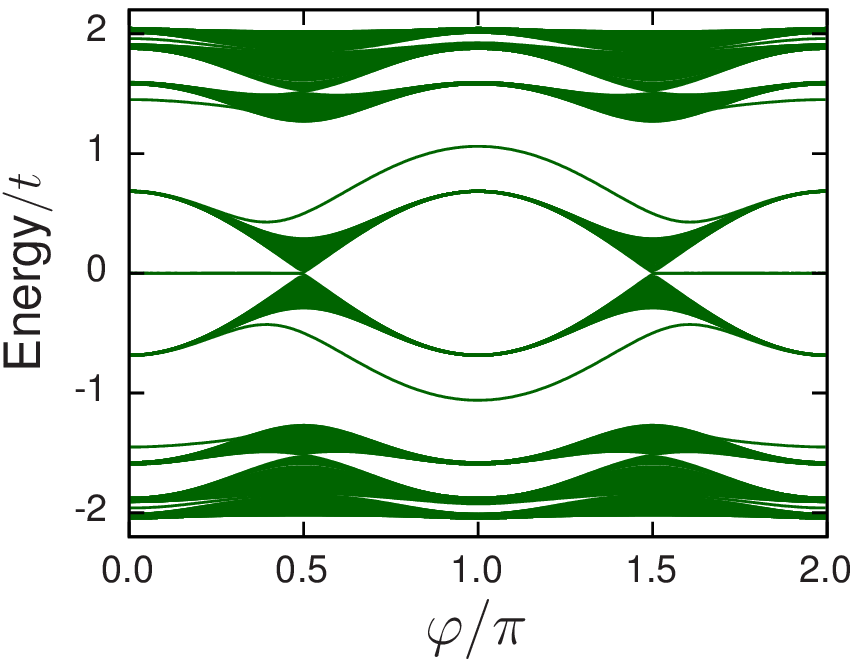} 
  \includegraphics[width=5.8cm]{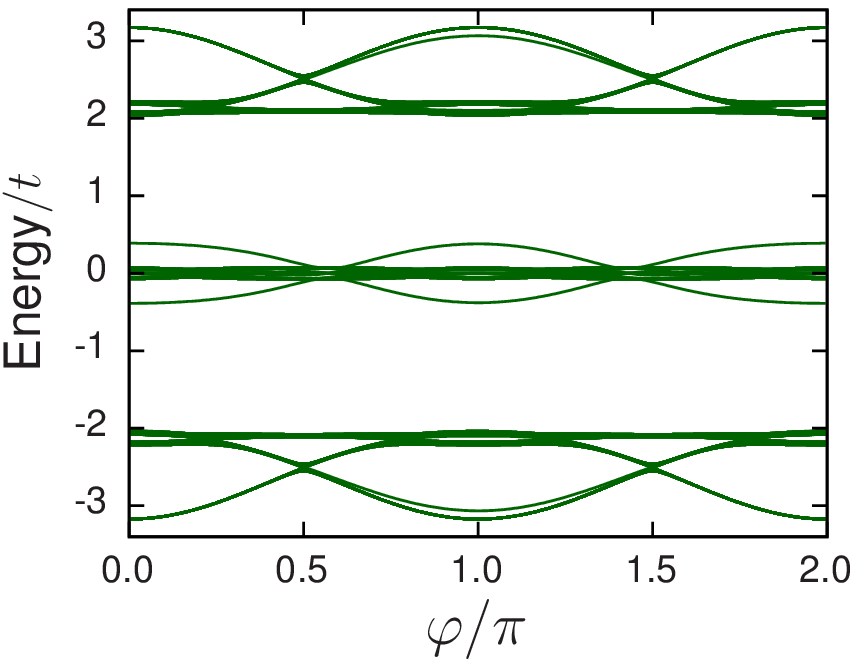} 
  \caption{Energy spectrum as a function of $\varphi$ for
    $\lambda=0.5$ and $\varphi_\delta = 0$, with $\delta = 0, 0.5,
    1.5$ (from left to right). Gaps open up under the incommensurate
    perturbation, but eventually states are pushed to the low-energy
    region.}  \label{fig:spectr01}
\end{figure*}

The purely commensurate off-diagonal model with $b=1/2$ shows
degenerate pairs of zero-energy topological states in the phase region
$|\varphi| < \pi/2$ (and equivalent regions displaced by $2\pi$)
\cite{topolSarma}. These states can be seen in the first plot of
Fig.~\ref{fig:spectr01}. Zero-energy topological states may be
associated with Majorana fermions \cite{majorana}. Such particles are
their own antiparticles, i.e., creation and annihilation operators are
equal. They can be defined as linear combinations of creation and
annihilation operators for real fermions, which is possible in a
particle-hole symmetric situation. Kitaev \cite{kitaev} used these
operators in a simple mean-field model of a 1D superconductor with
$p$-wave nearest-neighbor pairing. It defines a chain in which
alternate pairs of sites are coupled, leaving two unpaired Majorana
fermions at the ends. A real 1D superconductor, with spin-$1/2$
electrons, would not be time-reversal invariant in this case, and
should belong to the topology class D
\cite{Schnyderetal2008}. However, the model as originally proposed,
with spinless fermions, is both time-reversal invariant and
particle-hole symmetric, which implies chiral (or sublattice)
symmetry. So, the spinless Kitaev-model is classified into the BDI
topology class \cite{Schnyderetal2008}. This is the same topology of a
tight-binding chain with alternating hopping integrals $t_1, t_2$,
related to a model of polyacetylene \cite{polyacetylene}. It turns out
to be also the topology of the purely commensurate generalized AA
model with $b = 1/2$ in the phase region where topological states
exist, where we have $t_1 = t(1 - \lambda\cos\varphi)$ and $t_2 = t (1
+ \lambda\cos\varphi)$.

As implied by the above discussion, the observed topological states
are \emph{edge states}, hence only appearing for open boundary
conditions. One should notice that, in contrast to what happens in the
QHE, edge states in a 1D system are localized. Then, they may in
principle survive after the system undergoes Anderson-like
localization. We will show that it actually happens in the case
off-diagonal disorder. In contrast, a diagonal AA potential displaces
the edge-states away from zero energy for any $\Delta \ne 0$, which is
consistent with the breakdown of chiral symmetry by local disorder.

\begin{figure}
    \includegraphics[width=8cm]{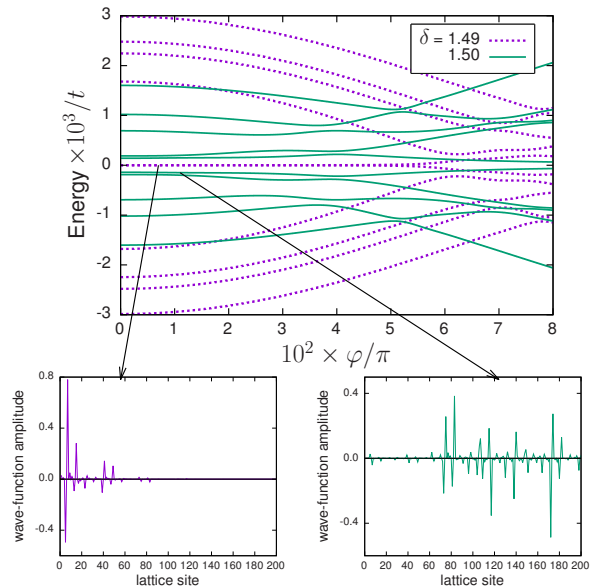}
    \caption{Expanded view (notice the scale factors) of the spectra
      for $\lambda=0.5$ and two values of $\delta$ around the critical
      value for which topological states are
      suppressed. Representative wavefunctions (lower plots) show the
      edge character of zero-energy states (left), absent when they
      split off (right).}  \label{fig:spectr02}
\end{figure}

In Fig.~\ref{fig:spectr01} we plot the energy eigenvalues as functions
of $\varphi$, for $\lambda=0.5$ and fixed $\varphi_\delta = 0$,
showing the trend of spectrum evolution as the disorder strength
$\delta$ increases. Starting from the purely commensurate case (first
plot), we observe that: (\emph{i}) gaps open up, and the bands are
substantially reshaped when disorder is turned on (middle plot), but
the topological states have not changed; (\emph{iii}) at $\delta=1.5$
(last plot), band states have been pushed to the middle of the gap,
and the topological states are no longer visible. Thus, we find that
the topological states survive the localization transition, but they
eventually disappear at a new critical value $\bar\delta_c >
\delta_c^{}$.  This is better seen in an expanded view of the
low-energy region presented in Fig.~\ref{fig:spectr02}. It highlights
the states with energy close to zero near $\varphi=0$, showing that in
this case ($\lambda=0.5$) the zero-energy degeneracy is lifted for
$\bar\delta_c \simeq 1.5$, while at a slightly smaller value of
$\delta$ the topological states are still clearly visible. The lower
panels in Fig.~\ref{fig:spectr02} are plots of wave-function
amplitudes corresponding to the indicated eigenvalues. At
$\bar\delta_c^{}$ the states no longer have the edge character still
noticeable for $\delta=1.49$.

\begin{figure}
    \includegraphics[width=8.5cm]{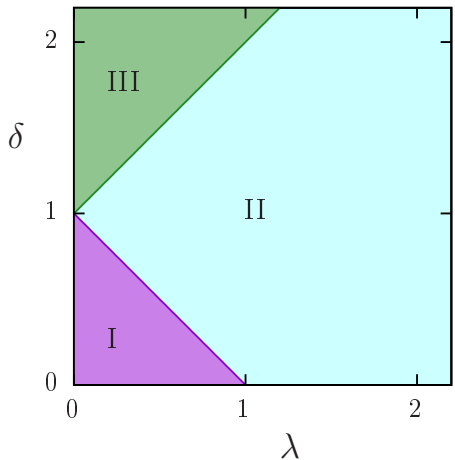}
    \caption{Phase diagram of model (\ref{eq:HAA}) with
      $\Delta=0$. The phases are (I) conductor with Majorana states,
      (II) Anderson insulator with Majorana states, and (III) Anderson
      insulator without Majorana states.} \label{fig:phdiag}
\end{figure}

As observed for $\delta_c^{}$ in the localization transition, the
critical value $\bar\delta_c$ also varies with
$\lambda$. Systematically studying this variation, we found that it
also follows a simple linear relationship, which in this case is
$\bar\delta_c = 1 + \lambda$. Based on this, we constructed a phase
diagram of the generalized AA model (without site-diagonal potential),
shown in Fig.~\ref{fig:phdiag}. It presents three distinct phases:
conductor with Majorana states (I), Anderson insulator
  with Majorana states (II), and Anderson insulator without
  Majorana states (III). By \emph{conductor} we mean a system in
which the bulk single-particle states are extended. 

The phase diagram of Fig.~\ref{fig:phdiag} is for null phases
($\varphi_\delta = \varphi = 0$). The role of $\varphi$ is easily
revealed. As we are restricting ourselves to the $b = 1/2$ model, the
$\lambda_i$ term in (\ref{eq:HAA}) is, in fact, $ \lambda \cos(\pi i +
\varphi) = \lambda \cos(\varphi)\, \cos(\pi i) = \lambda
\cos(\varphi)\,\cos(2\pi b i)$.  Therefore, the results for $\varphi
\ne 0$ can be directly obtained from the ones for $\varphi = 0$ by
just substituting $\lambda_\varphi \equiv \lambda \cos(\varphi)$ for
$\lambda$. Then, to obtain phase diagrams for nonzero values ​​of
$\varphi$ it suffices to rescale the horizontal axis in
Fig.~\ref{fig:phdiag} by a factor $1/\cos(\varphi)$. With this, the
value of $\lambda$ for which $\delta_c$ vanishes moves to the right,
and the angle between the two straight lines decreases. When $\varphi
= \pi/2$, the point $\lambda_\varphi = 1$ corresponds to $\lambda
\to \infty$, and the two lines coincide horizontally, implying that
the only transition occurs at $\delta_c = 1$, between a conductor and
an Anderson insulator. This happens at the ``Dirac points'' of the
purely commensurate spectrum (first plot in Fig.~\ref{fig:spectr01})
for which we have a single continuous band and no topological
states. For any finite $\lambda$, this is equivalent to a simple
lattice (uniform nearest-neighbor hopping), to which addition of an
incommensurate hopping modulation leads to localization at $\delta =
1$.

\begin{figure}
  \includegraphics[width=7.68cm]{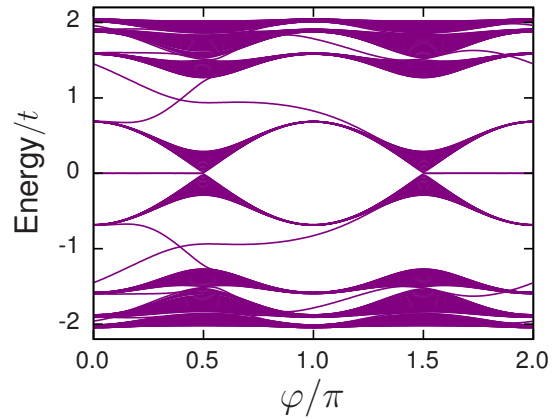}
  \caption{Energy spectrum of the generalized AA model with locked
    modulation phases $\varphi_\delta= \varphi$. In this example we
    have $\lambda =\delta= 0.5$. Notice that pairs of edge states
    connect neighboring bands across the gaps. Otherwise, the spectrum
    is very similar to the one for $\varphi_\delta=0$ shown in
    Fig.~\ref{fig:spectr01}.} \label{fig:specphiphi}.
\end{figure}

The effect of varying the phase $\varphi_\delta$ of the disorder term
is not as easy to describe in a general way since it is not possible
to absorb this phase into an effective amplitude. So far our analysis
has been restricted to $\varphi_\delta=0$. If we now lock the phases
of commensurate and incommensurate modulations ($\varphi_\delta =
\varphi$), the overall spectrum structure remains essentially as in
Fig.~\ref{fig:spectr01}, except for one important difference: the
bands never truly split off with increasing $\delta$, but remain
connected across the gaps by pairs of edge states. These are
reminiscent of edge states in topological insulators or the
QHE. However, Majorana states are also present in the region of small
$|\varphi_\delta|$ (mod $2\pi$). An example spectrum is presented in
Fig.~\ref{fig:specphiphi}. The low-energy region near $\varphi = 0$ in
the coupled-phases model is very similar to that presented in
Fig.~\ref{fig:spectr02}, indicating that the survival and subsequent
disappearance of zero-energy states is not significantly changed by
phase locking in that region. However, we have preliminary evidence
that zero-energy states may continue to exist for other phase
values. This interesting possibility deserves further investigation.

The observed spectral differences indicate that the generalized
off-diagonal AA model with coupled phases and the one with the
incommensurate-term phase fixed at zero belong to different
topological classes.  The model with coupled phases has the same
topology as the purely commensurate one ($\lambda \ne 0, \delta = 0$),
since one spectrum can be ``deformed'' into the other, by variation of
the parameter $\delta$, without closing (or opening) gaps. This is not
the case with fixed $\varphi_\delta^{} = 0$, as shown in
Fig.~\ref{fig:spectr01}. The existence of different topologies can be
understood by the effective two-dimensionality of the AA model in its
correspondence to the HH model, as discussed in the Introduction. In
fact, the phase $\varphi$ is actually a degree of freedom since it is
proportional to the transverse momentum of the electrons in the 2D
model.

\section{Conclusions}

We investigated extensions of the one-dimensional Aubry-Andr\'e model
in which a commensurate modulation of the hopping amplitude gives rise
to topological states of zero energy, associated with Majorana
fermions. We focused on how an off-diagonal disorder, realized by an
incommensurate hopping modulation, affects the energy spectrum,
inducing Anderson-like localization. We found out that the
topological zero-energy states survive after the localization
transition up to a second critical value of the disorder strength. In
addition, we observed that topological properties depend on the
relationship between the phases of commensurate and incommensurate
modulations.

Real physical systems, ranging from polymers \cite{polyacetylene} and
solid-state nanostructures \cite{kitaev} to optically confined
cold-atom lattices \cite{Roati08,Billy08,Nakajima2016} and light
propagation in waveguide arrays \cite{pugatch2,KrausZilb01}, can be
described by the kind of models studied here.  The fact that
quasi-periodic potentials can be realized in these systems implies
that experimental investigations of the interplay between topology and
disorder that we addressed here can be carried on.

\acknowledgments

We acknowledge support from the Brazilian agency Conselho Nacional
de Desenvolvimento Cient\'ifico e Tecnol\'ogico.


\begin{thebibliography}{31}%
\makeatletter
\providecommand \@ifxundefined [1]{%
 \@ifx{#1\undefined}
}%
\providecommand \@ifnum [1]{%
 \ifnum #1\expandafter \@firstoftwo
 \else \expandafter \@secondoftwo
 \fi
}%
\providecommand \@ifx [1]{%
 \ifx #1\expandafter \@firstoftwo
 \else \expandafter \@secondoftwo
 \fi
}%
\providecommand \natexlab [1]{#1}%
\providecommand \enquote  [1]{``#1''}%
\providecommand \bibnamefont  [1]{#1}%
\providecommand \bibfnamefont [1]{#1}%
\providecommand \citenamefont [1]{#1}%
\providecommand \href@noop [0]{\@secondoftwo}%
\providecommand \href [0]{\begingroup \@sanitize@url \@href}%
\providecommand \@href[1]{\@@startlink{#1}\@@href}%
\providecommand \@@href[1]{\endgroup#1\@@endlink}%
\providecommand \@sanitize@url [0]{\catcode `\\12\catcode `\$12\catcode
  `\&12\catcode `\#12\catcode `\^12\catcode `\_12\catcode `\%12\relax}%
\providecommand \@@startlink[1]{}%
\providecommand \@@endlink[0]{}%
\providecommand \url  [0]{\begingroup\@sanitize@url \@url }%
\providecommand \@url [1]{\endgroup\@href {#1}{\urlprefix }}%
\providecommand \urlprefix  [0]{URL }%
\providecommand \Eprint [0]{\href }%
\providecommand \doibase [0]{http://dx.doi.org/}%
\providecommand \selectlanguage [0]{\@gobble}%
\providecommand \bibinfo  [0]{\@secondoftwo}%
\providecommand \bibfield  [0]{\@secondoftwo}%
\providecommand \translation [1]{[#1]}%
\providecommand \BibitemOpen [0]{}%
\providecommand \bibitemStop [0]{}%
\providecommand \bibitemNoStop [0]{.\EOS\space}%
\providecommand \EOS [0]{\spacefactor3000\relax}%
\providecommand \BibitemShut  [1]{\csname bibitem#1\endcsname}%
\let\auto@bib@innerbib\@empty
\bibitem [{\citenamefont {Billy}\ \emph {et~al.}(2008)\citenamefont {Billy},
  \citenamefont {Josse}, \citenamefont {Zuo}, \citenamefont {Bernard},
  \citenamefont {Hambrecht}, \citenamefont {Lugan}, \citenamefont {Cl\'ement},
  \citenamefont {Sanchez-Palencia}, \citenamefont {Bouyer},\ and\ \citenamefont
  {Aspect}}]{Billy08}%
  \BibitemOpen
  \bibfield  {author} {\bibinfo {author} {\bibfnamefont {J.}~\bibnamefont
  {Billy}}, \bibinfo {author} {\bibfnamefont {V.}~\bibnamefont {Josse}},
  \bibinfo {author} {\bibfnamefont {Z.}~\bibnamefont {Zuo}}, \bibinfo {author}
  {\bibfnamefont {A.}~\bibnamefont {Bernard}}, \bibinfo {author} {\bibfnamefont
  {B.}~\bibnamefont {Hambrecht}}, \bibinfo {author} {\bibfnamefont
  {P.}~\bibnamefont {Lugan}}, \bibinfo {author} {\bibfnamefont
  {D.}~\bibnamefont {Cl\'ement}}, \bibinfo {author} {\bibfnamefont
  {L.}~\bibnamefont {Sanchez-Palencia}}, \bibinfo {author} {\bibfnamefont
  {P.}~\bibnamefont {Bouyer}}, \ and\ \bibinfo {author} {\bibfnamefont
  {A.}~\bibnamefont {Aspect}},\ }\href@noop {} {\bibfield  {journal} {\bibinfo
  {journal} {Nature}\ }\textbf {\bibinfo {volume} {453}},\ \bibinfo {pages}
  {891} (\bibinfo {year} {2008})}\BibitemShut {NoStop}%
\bibitem [{\citenamefont {Roati}\ \emph {et~al.}(2008)\citenamefont {Roati},
  \citenamefont {D'Errico}, \citenamefont {Fallani}, \citenamefont {Fattori},
  \citenamefont {Fort}, \citenamefont {Zaccanti}, \citenamefont {Modugno},
  \citenamefont {Modugno},\ and\ \citenamefont {Inguscio}}]{Roati08}%
  \BibitemOpen
  \bibfield  {author} {\bibinfo {author} {\bibfnamefont {G.}~\bibnamefont
  {Roati}}, \bibinfo {author} {\bibfnamefont {C.}~\bibnamefont {D'Errico}},
  \bibinfo {author} {\bibfnamefont {L.}~\bibnamefont {Fallani}}, \bibinfo
  {author} {\bibfnamefont {M.}~\bibnamefont {Fattori}}, \bibinfo {author}
  {\bibfnamefont {C.}~\bibnamefont {Fort}}, \bibinfo {author} {\bibfnamefont
  {M.}~\bibnamefont {Zaccanti}}, \bibinfo {author} {\bibfnamefont
  {G.}~\bibnamefont {Modugno}}, \bibinfo {author} {\bibfnamefont
  {M.}~\bibnamefont {Modugno}}, \ and\ \bibinfo {author} {\bibfnamefont
  {M.}~\bibnamefont {Inguscio}},\ }\href@noop {} {\bibfield  {journal}
  {\bibinfo  {journal} {Nature}\ }\textbf {\bibinfo {volume} {453}},\ \bibinfo
  {pages} {895} (\bibinfo {year} {2008})}\BibitemShut {NoStop}%
\bibitem [{\citenamefont {Lahini}\ \emph {et~al.}(2009)\citenamefont {Lahini},
  \citenamefont {Pugatch}, \citenamefont {Pozzi}, \citenamefont {Sorel},
  \citenamefont {Morandotti}, \citenamefont {Davidson},\ and\ \citenamefont
  {Silberberg}}]{pugatch2}%
  \BibitemOpen
  \bibfield  {author} {\bibinfo {author} {\bibfnamefont {Y.}~\bibnamefont
  {Lahini}}, \bibinfo {author} {\bibfnamefont {R.}~\bibnamefont {Pugatch}},
  \bibinfo {author} {\bibfnamefont {F.}~\bibnamefont {Pozzi}}, \bibinfo
  {author} {\bibfnamefont {M.}~\bibnamefont {Sorel}}, \bibinfo {author}
  {\bibfnamefont {R.}~\bibnamefont {Morandotti}}, \bibinfo {author}
  {\bibfnamefont {N.}~\bibnamefont {Davidson}}, \ and\ \bibinfo {author}
  {\bibfnamefont {Y.}~\bibnamefont {Silberberg}},\ }\href@noop {} {\bibfield
  {journal} {\bibinfo  {journal} {Phys. Rev. Lett.}\ }\textbf {\bibinfo
  {volume} {103}},\ \bibinfo {pages} {013901} (\bibinfo {year}
  {2009})}\BibitemShut {NoStop}%
\bibitem [{\citenamefont {Anderson}(1958)}]{Anderson}%
  \BibitemOpen
  \bibfield  {author} {\bibinfo {author} {\bibfnamefont {P.~W.}\ \bibnamefont
  {Anderson}},\ }\href@noop {} {\bibfield  {journal} {\bibinfo  {journal}
  {Phys. Rev.}\ }\textbf {\bibinfo {volume} {109}},\ \bibinfo {pages} {1492}
  (\bibinfo {year} {1958})}\BibitemShut {NoStop}%
\bibitem [{\citenamefont {Das~Sarma}\ \emph {et~al.}(1988)\citenamefont
  {Das~Sarma}, \citenamefont {He},\ and\ \citenamefont {Xie}}]{AndLocSarma01}%
  \BibitemOpen
  \bibfield  {author} {\bibinfo {author} {\bibfnamefont {S.}~\bibnamefont
  {Das~Sarma}}, \bibinfo {author} {\bibfnamefont {S.}~\bibnamefont {He}}, \
  and\ \bibinfo {author} {\bibfnamefont {X.~C.}\ \bibnamefont {Xie}},\ }\href
  {\doibase 10.1103/PhysRevLett.61.2144} {\bibfield  {journal} {\bibinfo
  {journal} {Phys. Rev. Lett.}\ }\textbf {\bibinfo {volume} {61}},\ \bibinfo
  {pages} {2144} (\bibinfo {year} {1988})}\BibitemShut {NoStop}%
\bibitem [{\citenamefont {Das~Sarma}\ \emph {et~al.}(1990)\citenamefont
  {Das~Sarma}, \citenamefont {He},\ and\ \citenamefont {Xie}}]{AndLocSarma02}%
  \BibitemOpen
  \bibfield  {author} {\bibinfo {author} {\bibfnamefont {S.}~\bibnamefont
  {Das~Sarma}}, \bibinfo {author} {\bibfnamefont {S.}~\bibnamefont {He}}, \
  and\ \bibinfo {author} {\bibfnamefont {X.~C.}\ \bibnamefont {Xie}},\ }\href
  {\doibase 10.1103/PhysRevB.41.5544} {\bibfield  {journal} {\bibinfo
  {journal} {Phys. Rev. B}\ }\textbf {\bibinfo {volume} {41}},\ \bibinfo
  {pages} {5544} (\bibinfo {year} {1990})}\BibitemShut {NoStop}%
\bibitem [{\citenamefont {Biddle}\ \emph {et~al.}(2009)\citenamefont {Biddle},
  \citenamefont {Wang}, \citenamefont {Priour},\ and\ \citenamefont
  {Das~Sarma}}]{AndLocSarma03}%
  \BibitemOpen
  \bibfield  {author} {\bibinfo {author} {\bibfnamefont {J.}~\bibnamefont
  {Biddle}}, \bibinfo {author} {\bibfnamefont {B.}~\bibnamefont {Wang}},
  \bibinfo {author} {\bibfnamefont {D.~J.}\ \bibnamefont {Priour}}, \ and\
  \bibinfo {author} {\bibfnamefont {S.}~\bibnamefont {Das~Sarma}},\ }\href
  {\doibase 10.1103/PhysRevA.80.021603} {\bibfield  {journal} {\bibinfo
  {journal} {Phys. Rev. A}\ }\textbf {\bibinfo {volume} {80}},\ \bibinfo
  {pages} {021603} (\bibinfo {year} {2009})}\BibitemShut {NoStop}%
\bibitem [{\citenamefont {Biddle}\ and\ \citenamefont
  {Das~Sarma}(2010)}]{AndLocSarma04}%
  \BibitemOpen
  \bibfield  {author} {\bibinfo {author} {\bibfnamefont {J.}~\bibnamefont
  {Biddle}}\ and\ \bibinfo {author} {\bibfnamefont {S.}~\bibnamefont
  {Das~Sarma}},\ }\href {\doibase 10.1103/PhysRevLett.104.070601} {\bibfield
  {journal} {\bibinfo  {journal} {Phys. Rev. Lett.}\ }\textbf {\bibinfo
  {volume} {104}},\ \bibinfo {pages} {070601} (\bibinfo {year}
  {2010})}\BibitemShut {NoStop}%
\bibitem [{\citenamefont {Cestari}\ \emph {et~al.}(2010)\citenamefont
  {Cestari}, \citenamefont {Foerster},\ and\ \citenamefont
  {Gusm\~ao}}]{cestari10}%
  \BibitemOpen
  \bibfield  {author} {\bibinfo {author} {\bibfnamefont {J.~C.~C.}\
  \bibnamefont {Cestari}}, \bibinfo {author} {\bibfnamefont {A.}~\bibnamefont
  {Foerster}}, \ and\ \bibinfo {author} {\bibfnamefont {M.~A.}\ \bibnamefont
  {Gusm\~ao}},\ }\href {\doibase 10.1103/PhysRevA.82.063634} {\bibfield
  {journal} {\bibinfo  {journal} {Phys. Rev. A}\ }\textbf {\bibinfo {volume}
  {82}},\ \bibinfo {pages} {063634} (\bibinfo {year} {2010})}\BibitemShut
  {NoStop}%
\bibitem [{\citenamefont {Cestari}\ \emph {et~al.}(2011)\citenamefont
  {Cestari}, \citenamefont {Foerster}, \citenamefont {Gusm{\~a}o},\ and\
  \citenamefont {Continentino}}]{cestari12}%
  \BibitemOpen
  \bibfield  {author} {\bibinfo {author} {\bibfnamefont {J.~C.~C.}\
  \bibnamefont {Cestari}}, \bibinfo {author} {\bibfnamefont {A.}~\bibnamefont
  {Foerster}}, \bibinfo {author} {\bibfnamefont {M.~A.}\ \bibnamefont
  {Gusm{\~a}o}}, \ and\ \bibinfo {author} {\bibfnamefont {M.}~\bibnamefont
  {Continentino}},\ }\href@noop {} {\bibfield  {journal} {\bibinfo  {journal}
  {Physical Review A}\ }\textbf {\bibinfo {volume} {84}},\ \bibinfo {pages}
  {055601} (\bibinfo {year} {2011})}\BibitemShut {NoStop}%
\bibitem [{\citenamefont {Aubry}\ and\ \citenamefont {Andr\'e}(1980)}]{AA}%
  \BibitemOpen
  \bibfield  {author} {\bibinfo {author} {\bibfnamefont {S.}~\bibnamefont
  {Aubry}}\ and\ \bibinfo {author} {\bibfnamefont {G.}~\bibnamefont
  {Andr\'e}},\ }\href@noop {} {\bibfield  {journal} {\bibinfo  {journal} {Ann.
  Israel Phys. Soc.}\ }\textbf {\bibinfo {volume} {3}},\ \bibinfo {pages} {133}
  (\bibinfo {year} {1980})}\BibitemShut {NoStop}%
\bibitem [{\citenamefont {Harper}(1955)}]{harper}%
  \BibitemOpen
  \bibfield  {author} {\bibinfo {author} {\bibfnamefont {P.~G.}\ \bibnamefont
  {Harper}},\ }\href@noop {} {\bibfield  {journal} {\bibinfo  {journal} {Proc.
  Phys. Soc. Lond. Sect. A}\ }\textbf {\bibinfo {volume} {68}},\ \bibinfo
  {pages} {874} (\bibinfo {year} {1955})}\BibitemShut {NoStop}%
\bibitem [{\citenamefont {Hofstadter}(1976)}]{Hofstadter}%
  \BibitemOpen
  \bibfield  {author} {\bibinfo {author} {\bibfnamefont {D.~R.}\ \bibnamefont
  {Hofstadter}},\ }\href {\doibase 10.1103/PhysRevB.14.2239} {\bibfield
  {journal} {\bibinfo  {journal} {Phys. Rev. B}\ }\textbf {\bibinfo {volume}
  {14}},\ \bibinfo {pages} {2239} (\bibinfo {year} {1976})}\BibitemShut
  {NoStop}%
\bibitem [{\citenamefont {Lang}\ \emph {et~al.}(2012)\citenamefont {Lang},
  \citenamefont {Cai},\ and\ \citenamefont {Chen}}]{topo1}%
  \BibitemOpen
  \bibfield  {author} {\bibinfo {author} {\bibfnamefont {L.-J.}\ \bibnamefont
  {Lang}}, \bibinfo {author} {\bibfnamefont {X.}~\bibnamefont {Cai}}, \ and\
  \bibinfo {author} {\bibfnamefont {S.}~\bibnamefont {Chen}},\ }\href@noop {}
  {\bibfield  {journal} {\bibinfo  {journal} {Physical Review Letters}\
  }\textbf {\bibinfo {volume} {108}},\ \bibinfo {pages} {220401} (\bibinfo
  {year} {2012})}\BibitemShut {NoStop}%
\bibitem [{\citenamefont {Kraus}\ \emph {et~al.}(2012)\citenamefont {Kraus},
  \citenamefont {Lahini}, \citenamefont {Ringel}, \citenamefont {Verbin},\ and\
  \citenamefont {Zilberberg}}]{KrausZilb01}%
  \BibitemOpen
  \bibfield  {author} {\bibinfo {author} {\bibfnamefont {Y.~E.}\ \bibnamefont
  {Kraus}}, \bibinfo {author} {\bibfnamefont {Y.}~\bibnamefont {Lahini}},
  \bibinfo {author} {\bibfnamefont {Z.}~\bibnamefont {Ringel}}, \bibinfo
  {author} {\bibfnamefont {M.}~\bibnamefont {Verbin}}, \ and\ \bibinfo {author}
  {\bibfnamefont {O.}~\bibnamefont {Zilberberg}},\ }\href@noop {} {\bibfield
  {journal} {\bibinfo  {journal} {Physical Review Letters}\ }\textbf {\bibinfo
  {volume} {109}},\ \bibinfo {pages} {106402} (\bibinfo {year}
  {2012})}\BibitemShut {NoStop}%
\bibitem [{\citenamefont {Kraus}\ and\ \citenamefont
  {Zilberberg}(2012)}]{KrausZilb02}%
  \BibitemOpen
  \bibfield  {author} {\bibinfo {author} {\bibfnamefont {Y.~E.}\ \bibnamefont
  {Kraus}}\ and\ \bibinfo {author} {\bibfnamefont {O.}~\bibnamefont
  {Zilberberg}},\ }\href@noop {} {\bibfield  {journal} {\bibinfo  {journal}
  {Physical Review Letters}\ }\textbf {\bibinfo {volume} {109}},\ \bibinfo
  {pages} {116404} (\bibinfo {year} {2012})}\BibitemShut {NoStop}%
\bibitem [{\citenamefont {Hasan}\ and\ \citenamefont
  {Kane}(2010)}]{topo_insul}%
  \BibitemOpen
  \bibfield  {author} {\bibinfo {author} {\bibfnamefont {M.~Z.}\ \bibnamefont
  {Hasan}}\ and\ \bibinfo {author} {\bibfnamefont {C.~L.}\ \bibnamefont
  {Kane}},\ }\href@noop {} {\bibfield  {journal} {\bibinfo  {journal} {Reviews
  of Modern Physics}\ }\textbf {\bibinfo {volume} {82}},\ \bibinfo {pages}
  {3045} (\bibinfo {year} {2010})}\BibitemShut {NoStop}%
\bibitem [{\citenamefont {Leijnse}\ and\ \citenamefont
  {Flensberg}(2012)}]{majorana}%
  \BibitemOpen
  \bibfield  {author} {\bibinfo {author} {\bibfnamefont {M.}~\bibnamefont
  {Leijnse}}\ and\ \bibinfo {author} {\bibfnamefont {K.}~\bibnamefont
  {Flensberg}},\ }\href@noop {} {\bibfield  {journal} {\bibinfo  {journal}
  {Semiconductor Science and Technology}\ }\textbf {\bibinfo {volume} {27}},\
  \bibinfo {pages} {124003} (\bibinfo {year} {2012})}\BibitemShut {NoStop}%
\bibitem [{\citenamefont {Laughlin}(1981)}]{QHE01}%
  \BibitemOpen
  \bibfield  {author} {\bibinfo {author} {\bibfnamefont {R.~B.}\ \bibnamefont
  {Laughlin}},\ }\href {\doibase 10.1103/PhysRevB.23.5632} {\bibfield
  {journal} {\bibinfo  {journal} {Phys. Rev. B}\ }\textbf {\bibinfo {volume}
  {23}},\ \bibinfo {pages} {5632} (\bibinfo {year} {1981})}\BibitemShut
  {NoStop}%
\bibitem [{\citenamefont {Thouless}\ \emph {et~al.}(1982)\citenamefont
  {Thouless}, \citenamefont {Kohmoto}, \citenamefont {Nightingale},\ and\
  \citenamefont {den Nijs}}]{QHE02}%
  \BibitemOpen
  \bibfield  {author} {\bibinfo {author} {\bibfnamefont {D.~J.}\ \bibnamefont
  {Thouless}}, \bibinfo {author} {\bibfnamefont {M.}~\bibnamefont {Kohmoto}},
  \bibinfo {author} {\bibfnamefont {M.~P.}\ \bibnamefont {Nightingale}}, \ and\
  \bibinfo {author} {\bibfnamefont {M.}~\bibnamefont {den Nijs}},\ }\href
  {\doibase 10.1103/PhysRevLett.49.405} {\bibfield  {journal} {\bibinfo
  {journal} {Phys. Rev. Lett.}\ }\textbf {\bibinfo {volume} {49}},\ \bibinfo
  {pages} {405} (\bibinfo {year} {1982})}\BibitemShut {NoStop}%
\bibitem [{\citenamefont {Hatsugai}(1993)}]{QHE03}%
  \BibitemOpen
  \bibfield  {author} {\bibinfo {author} {\bibfnamefont {Y.}~\bibnamefont
  {Hatsugai}},\ }\href {\doibase 10.1103/PhysRevLett.71.3697} {\bibfield
  {journal} {\bibinfo  {journal} {Phys. Rev. Lett.}\ }\textbf {\bibinfo
  {volume} {71}},\ \bibinfo {pages} {3697} (\bibinfo {year}
  {1993})}\BibitemShut {NoStop}%
\bibitem [{\citenamefont {Liu}\ \emph {et~al.}(2015)\citenamefont {Liu},
  \citenamefont {Ghosh},\ and\ \citenamefont {Chong}}]{Liuetal2015}%
  \BibitemOpen
  \bibfield  {author} {\bibinfo {author} {\bibfnamefont {F.}~\bibnamefont
  {Liu}}, \bibinfo {author} {\bibfnamefont {S.}~\bibnamefont {Ghosh}}, \ and\
  \bibinfo {author} {\bibfnamefont {Y.~D.}\ \bibnamefont {Chong}},\ }\href
  {\doibase 10.1103/PhysRevB.91.014108} {\bibfield  {journal} {\bibinfo
  {journal} {Phys. Rev. B}\ }\textbf {\bibinfo {volume} {91}},\ \bibinfo
  {pages} {014108} (\bibinfo {year} {2015})}\BibitemShut {NoStop}%
\bibitem [{\citenamefont {Ganeshan}\ \emph {et~al.}(2013)\citenamefont
  {Ganeshan}, \citenamefont {Sun},\ and\ \citenamefont
  {Das~Sarma}}]{topolSarma}%
  \BibitemOpen
  \bibfield  {author} {\bibinfo {author} {\bibfnamefont {S.}~\bibnamefont
  {Ganeshan}}, \bibinfo {author} {\bibfnamefont {K.}~\bibnamefont {Sun}}, \
  and\ \bibinfo {author} {\bibfnamefont {S.}~\bibnamefont {Das~Sarma}},\ }\href
  {\doibase 10.1103/PhysRevLett.110.180403} {\bibfield  {journal} {\bibinfo
  {journal} {Phys. Rev. Lett.}\ }\textbf {\bibinfo {volume} {110}},\ \bibinfo
  {pages} {180403} (\bibinfo {year} {2013})}\BibitemShut {NoStop}%
\bibitem [{\citenamefont {Roth}\ and\ \citenamefont {Burnett}(2003)}]{roth}%
  \BibitemOpen
  \bibfield  {author} {\bibinfo {author} {\bibfnamefont {R.}~\bibnamefont
  {Roth}}\ and\ \bibinfo {author} {\bibfnamefont {K.}~\bibnamefont {Burnett}},\
  }\href@noop {} {\bibfield  {journal} {\bibinfo  {journal} {Phys. Rev. A}\
  }\textbf {\bibinfo {volume} {68}},\ \bibinfo {pages} {023604} (\bibinfo
  {year} {2003})}\BibitemShut {NoStop}%
\bibitem [{\citenamefont {Nielsen}\ and\ \citenamefont
  {Chuang}(2000)}]{nielsen}%
  \BibitemOpen
  \bibfield  {author} {\bibinfo {author} {\bibfnamefont {M.~A.}\ \bibnamefont
  {Nielsen}}\ and\ \bibinfo {author} {\bibfnamefont {I.~L.}\ \bibnamefont
  {Chuang}},\ }\href@noop {} {\emph {\bibinfo {title} {Quantum Computation and
  Quantum Information}}}\ (\bibinfo  {publisher} {Cambridge University Press},\
  \bibinfo {year} {2000})\BibitemShut {NoStop}%
\bibitem [{\citenamefont {Kohmoto}(1983)}]{kohmoto1}%
  \BibitemOpen
  \bibfield  {author} {\bibinfo {author} {\bibfnamefont {M.}~\bibnamefont
  {Kohmoto}},\ }\href@noop {} {\bibfield  {journal} {\bibinfo  {journal} {Phys.
  Rev. Lett.}\ }\textbf {\bibinfo {volume} {51}},\ \bibinfo {pages} {1198}
  (\bibinfo {year} {1983})}\BibitemShut {NoStop}%
\bibitem [{\citenamefont {Fisher}\ \emph {et~al.}(1973)\citenamefont {Fisher},
  \citenamefont {Barber},\ and\ \citenamefont {Jasnow}}]{fisherpai}%
  \BibitemOpen
  \bibfield  {author} {\bibinfo {author} {\bibfnamefont {M.~E.}\ \bibnamefont
  {Fisher}}, \bibinfo {author} {\bibfnamefont {M.~N.}\ \bibnamefont {Barber}},
  \ and\ \bibinfo {author} {\bibfnamefont {D.}~\bibnamefont {Jasnow}},\ }\href
  {\doibase 10.1103/PhysRevA.8.1111} {\bibfield  {journal} {\bibinfo  {journal}
  {Phys. Rev. A}\ }\textbf {\bibinfo {volume} {8}},\ \bibinfo {pages} {1111}
  (\bibinfo {year} {1973})}\BibitemShut {NoStop}%
\bibitem [{\citenamefont {Kitaev}(2001)}]{kitaev}%
  \BibitemOpen
  \bibfield  {author} {\bibinfo {author} {\bibfnamefont {A.~Y.}\ \bibnamefont
  {Kitaev}},\ }\href@noop {} {\bibfield  {journal} {\bibinfo  {journal}
  {Physics-Uspekhi}\ }\textbf {\bibinfo {volume} {44}},\ \bibinfo {pages} {131}
  (\bibinfo {year} {2001})}\BibitemShut {NoStop}%
\bibitem [{\citenamefont {Schnyder}\ \emph {et~al.}(2008)\citenamefont
  {Schnyder}, \citenamefont {Ryu}, \citenamefont {Furusaki},\ and\
  \citenamefont {Ludwig}}]{Schnyderetal2008}%
  \BibitemOpen
  \bibfield  {author} {\bibinfo {author} {\bibfnamefont {A.~P.}\ \bibnamefont
  {Schnyder}}, \bibinfo {author} {\bibfnamefont {S.}~\bibnamefont {Ryu}},
  \bibinfo {author} {\bibfnamefont {A.}~\bibnamefont {Furusaki}}, \ and\
  \bibinfo {author} {\bibfnamefont {A.~W.~W.}\ \bibnamefont {Ludwig}},\ }\href
  {\doibase 10.1103/PhysRevB.78.195125} {\bibfield  {journal} {\bibinfo
  {journal} {Phys. Rev. B}\ }\textbf {\bibinfo {volume} {78}},\ \bibinfo
  {pages} {195125} (\bibinfo {year} {2008})}\BibitemShut {NoStop}%
\bibitem [{\citenamefont {Su}\ \emph {et~al.}(1979)\citenamefont {Su},
  \citenamefont {Schrieffer},\ and\ \citenamefont {Heeger}}]{polyacetylene}%
  \BibitemOpen
  \bibfield  {author} {\bibinfo {author} {\bibfnamefont {W.~P.}\ \bibnamefont
  {Su}}, \bibinfo {author} {\bibfnamefont {J.~R.}\ \bibnamefont {Schrieffer}},
  \ and\ \bibinfo {author} {\bibfnamefont {A.~J.}\ \bibnamefont {Heeger}},\
  }\href {\doibase 10.1103/PhysRevLett.42.1698} {\bibfield  {journal} {\bibinfo
   {journal} {Phys. Rev. Lett.}\ }\textbf {\bibinfo {volume} {42}},\ \bibinfo
  {pages} {1698} (\bibinfo {year} {1979})}\BibitemShut {NoStop}%
\bibitem [{\citenamefont {Nakajima}\ \emph {et~al.}(2016)\citenamefont
    {Nakajima}, \citenamefont {Tomita}, \citenamefont {Taie},
    \citenamefont {Ichinose}, \citenamefont {Ozawa}, \citenamefont
    {Wang}, \citenamefont {Troyer},\ and\ \citenamefont
    {Takahashi}}]{Nakajima2016}%
  \BibitemOpen \bibfield {author} {\bibinfo {author} {\bibfnamefont
      {S.}~\bibnamefont {Nakajima}}, \bibinfo {author} {\bibfnamefont
      {T.}~\bibnamefont {Tomita}}, \bibinfo {author} {\bibfnamefont
      {S.}~\bibnamefont {Taie}}, \bibinfo {author} {\bibfnamefont
      {T.}~\bibnamefont {Ichinose}}, \bibinfo {author} {\bibfnamefont
      {H.}~\bibnamefont {Ozawa}}, \bibinfo {author} {\bibfnamefont
      {L.}~\bibnamefont {Wang}}, \bibinfo {author} {\bibfnamefont
      {M.}~\bibnamefont {Troyer}}, \ and\ \bibinfo {author}
    {\bibfnamefont {Y.}~\bibnamefont {Takahashi}},\ }\href
  {http://dx.doi.org/10.1038/nphys3622} {\bibfield {journal} {\bibinfo
      {journal} {Nat. Phys.}\ }\textbf {\bibinfo {volume} {12}},\
    \bibinfo {pages} {1698} (\bibinfo {year} {2016})}\BibitemShut
  {NoStop}%
\end{thebibliography}
\end{document}